# Transformed Naïve Ratio and Product Based Estimators for Estimating Population Mode in Simple Random Sampling

Sanjay Kumar[*] and Nirmal Tiwari
Department of Statistics, Central University of Rajasthan,
Bandarsindri, Kishangarh-305817, Ajmer, Rajasthan, India.,
[*]E-mail: sanjay.kumar@curaj.ac.in, Cell No.-(+91)-7597279362(Corresponding author)

## Abstract

In this paper, we propose a transformed naïve ratio and product based estimators using the characterizing scalar in presence of auxiliary information of the study variable for estimating the population mode following simple random sampling without replacement. The bias, mean square errors, relative efficiency, ratios of the exact values of mean square errors to the simulated mean square errors and confidence interval are studied for the performance of the proposed transformed naïve ratio type estimator with the certain natural population as well as artificially generated data sets. We have shown that proposed transformed naïve ratio based estimator is more efficient than the naïve estimator and naïve ratio estimator of the population mode.

## 4.1. Introduction

Survey statisticians generally wish to estimate population total, mean, median, quantile etc. for the study variable using auxiliary information. Auxiliary information helps to improve the efficiencies of estimators. Cochran (1940) was the first researcher who used auxiliary information to estimate the population total/mean while Kuk and Mak (1989) were the first researchers who used such auxiliary information to estimate the median of the study variable. However, there are many real life situations such as, manufacturing of shoes, manufacturing of ready-made garments and business forecasting etc., where to estimate population mode instead of population mean/total or median of the study variable, is more appropriate. There are extensive literatures on estimating the total/mean, median and quantile using auxiliary information. But works on estimating mode are not done enough. Various researchers such as Doodson (1917), Yasukawa (1926), Chernoff (1964), Grenander (1965), Dalenius (1965), Venter (1967), Robertson and Cryer (1974), Kendall and Stuart (1977) and Marlin (1983) worked on such problems. Sedory and Singh (2014) first suggested a ratio estimator using auxiliary information to estimate the population mode. In this chapter, we suggest transformed naive ratio-and product-type estimators using characterizing scalars to estimate the population mode of the study variable. The impact of suggested research work is expected to be useful for the scientists in the field of Sociology, Psychology, Demography or Actuarial Sciences, Business, Economics, Medicine etc., where the mode is routinely being used in practice. We proposed a transformed naïve ratio and product type estimators using some characterizing scalar for estimating population mode in presence of auxiliary information. The proposed work is motivated by Sedory and Singh (2014).

## 2. Notations

Let $y$ be the study variable and $x$ be the auxiliary variable, $i^{th}$ unit of population $i = 1,2 \ldots N$ is denoted by $y_i$ and $x_i$. Assume that, The sample of size $n$ drawn from the population of the size $N$ using the Simple random sampling without replacement (SRSWOR).

Let us define:

The sample means, sample variances and covariance terms are given as:

$$\bar{y} = \frac{1}{n}\sum_{i=1}^{n} y_i, \; \bar{x} = \frac{1}{n}\sum_{i=1}^{n} x_i, \; s_y^2 = \frac{1}{n-1}\sum_{i=1}^{n}(y_i - \bar{y})^2, \; s_x^2 = \frac{1}{n-1}\sum_{i=1}^{n}(x_i - \bar{x})^2,$$

$$S_{yx} = \frac{1}{n-1}\sum_{i=1}^{n}(y_i - \bar{y})(x_i - \bar{x})$$

The population means, population variances and covariance term given as:

$$\bar{Y} = \frac{1}{N}\sum_{i=1}^{N} y_i, \; \bar{X} = \frac{1}{N}\sum_{i=1}^{N} x_i, \; S_y^2 = \frac{1}{N-1}\sum_{i=1}^{N}(y_i - \bar{Y})^2, \; S_x^2 = \frac{1}{n-1}\sum_{i=1}^{N}(x_i - \bar{X})^2,$$

$$S_{yx} = \frac{1}{N-1}\sum_{i=1}^{N}(y_i - \bar{Y})(x_i - \bar{X})$$

Let $y_{(1)}, y_{(2)}, \ldots, y_{(n)}$ are the $y$ values of the sample and arranged in the order $y_{(1)} \leq y_{(2)} \leq \cdots \leq y_{(n)}$ and $p = \frac{t}{n}$ be the proportion of $y$ values. The median value $M_y$ is always greater than or equal to the $p$; the $p$ and $M_y$ are unknown which can be estimated. Kuk and Mak proposed Matrix of proportions $P_{ij}$ as:

|  | $y \leq M_y$ | $y \leq M_y$ | Total |
|---|---|---|---|
| $x \leq M_x$ | $P_{11}$ | $P_{21}$ | $P_{\cdot 1}$ |
| $x > M_x$ | $P_{12}$ | $P_{22}$ | $P_{\cdot 2}$ |
| Total | $P_{1\cdot}$ | $P_{2\cdot}$ | 1 |

The bivariate distribution of variable $(y,x)$ approaches a continues distribution when it is assumed that $N \to \infty$, with marginal densities $f_y(\cdot)$ and $f_x(\cdot)$ for the study as well as auxiliary variable $y$ and $x$ respectively. This assumption holds in particular under super population model framework, treating the values of $(y,x)$ in the population as a realization of $N$ independent observations from a continuous distribution. It is assumed that $f_y(\cdot)$ and $f_x(\cdot)$ are positive.

Under these assumptions, Gross(1980) has shown the sample median $\widehat{M}_y$ is the consistent and asymptotically normal with mean $M_y$ and asymptotic variance:

$$V(\widehat{M}_y) = \left(\frac{1-f}{4n}\right)\{f_y(M_y)\}^{-2} \qquad (2.1)$$

Where, $f = \frac{n}{N}$

Doodson(1917) proposed the empirical relationship of the mean, median and mode if the distribution is moderately asymptotic:

$$Mode \approx 3 Median - 2 Mean \qquad (2.2)$$

This is also known as Karl Pearson empirical relationship between mean, median and mode.

Let $\tilde{Y}$ and $\tilde{X}$ be the population mode value of study variable $y$ and auxiliary variable $x$ and is given by

$$\tilde{Y} \approx 3M_y - 2\bar{Y} \quad \text{and} \quad \tilde{X} \approx 3M_x - 2\bar{X} \qquad (2.3)$$

Where $M_y$ and $M_x$ is the population median of study and auxiliary variable.

The naïve estimator for the $\tilde{Y}$ and $\tilde{X}$ are the

$$\tilde{y} \approx 3\widehat{M}_y - 2\bar{y} \quad \text{and} \quad \tilde{x} \approx 3\widehat{M}_x - 2\bar{x} \qquad (2.4)$$

Where $\widehat{M}_y$ and $\widehat{M}_x$ is the sample median of study and auxiliary variable.

Let us define indicator function for the variable $y$ and variable $x$

$$I_{y_i} = \begin{cases} 1, & if\ y_i \leq M_y \\ 0, & otherwise \end{cases} \quad \text{and} \quad I_{x_i} = \begin{cases} 1, & if\ x_i \leq M_x \\ 0, & otherwise \end{cases} \qquad (2.5)$$

The variance and covariance of naïve estimators $\tilde{y}$ and $\tilde{x}$ may be approximated by

$$V(\tilde{y}) = \left\{\frac{1-f}{n}\right\}\left[\frac{9}{4}\{f_y(M_y)\}^{-2} + 4S_y^2 + 12S_{yM_y}\{f_y(M_y)\}^{-1}\right] \tag{2.6}$$

$$V(\tilde{x}) = \left\{\frac{1-f}{n}\right\}\left[\frac{9}{4}\{f_x(M_x)\}^{-2} + 4S_x^2 + 12S_{xM_x}\{f_x(M_x)\}^{-1}\right] \tag{2.7}$$

and

$Cov(\tilde{y}, \tilde{x}) =$
$\left(\frac{1-f}{n}\right)\left[9\{f_x(M_x)\,f_y(M_y)\}^{-1}(P_{11} - 0.25) + 6S_{yM_x}\{f_x(M_x)\}^{-1} + 6S_{xM_y}\{f_y(M_y)\}^{-1} + 4S_{xy}\right]$ (2.8)

where

$f = \frac{n}{N}$,

$S_{yM_y} = \frac{1}{N-1}\sum_{i=1}^{N}(y_i - \bar{Y})(I_{y_i} - 0.5)$, $S_{xM_x} = \frac{1}{N-1}\sum_{i=1}^{N}(x_i - \bar{X})(I_{x_i} - 0.5)$

$S_{yM_x} = \frac{1}{N-1}\sum_{i=1}^{N}(y_i - \bar{Y})(I_{y_i} - 0.5)$ and $S_{xM_y} = \frac{1}{N-1}\sum_{i=1}^{N}(x_i - \bar{X})(I_{y_i} - 0.05)$

For deriving the variance and covariance expressions of sample modes we have used the following main result of Kuk and Mak (1989), if $F_y$ be the cumulative distribution function of $y$, then the approximation of sample median is given by:

$$\widehat{M}_y = M_y + (0.5 - p_y)\{f_y(M_y)\}^{-1} + \cdots \tag{2.9}$$

Where $p_y$ is the proportion of $I_y$ values taking a value of 1. Further note that, we also verified that

$$(P_{11} - 0.25) = \frac{1}{N-1}\sum_{i=1}^{N}(I_{x_i} - 0.05)(I_{y_i} - 0.05) \tag{2.10}$$

Thus, we use the above result for finding covariance between two median matches with the result of Kuk and Mak(1989).

Let us define:

$\tilde{y} = \tilde{Y}(1 + \epsilon_0)$, $\tilde{x} = \tilde{X}(1 + \epsilon_1)$, such that $E(\epsilon_0) = E(\epsilon_1) = 0$

We are following simple random sampling without replacement (SRSWOR) method of sampling, then we have

$$E(\epsilon_0^2) = \frac{V(\tilde{y})}{\tilde{Y}^2}, \quad E(\epsilon_1^2) = \frac{V(\tilde{x})}{\tilde{X}^2} \text{ and } E(\epsilon_0\epsilon_1) = \frac{Cov(\tilde{y},\tilde{x})}{\tilde{Y}\tilde{X}}$$

The usual naïve ratio estimator due to Sedory and Singh (2014) of the population mode is given by

$$t_r = \tilde{y}\frac{\tilde{X}}{\tilde{x}} \qquad (2.11)$$

The expression of mean square errors of the naïve ratio estimator $t_r$'s is given by

$$MSE(t_r) \approx V(\tilde{y}) + \tilde{R}^2 V(\tilde{x}) - 2\tilde{R}\, Cov(\tilde{y},\tilde{x}) \qquad (2.12)$$

Where $\tilde{R} = \frac{\tilde{Y}}{\tilde{X}}$, is the ratio of the population modes of the study variable and the auxiliary variable.

The usual naïve product estimator following Sedory and Singh (2014) of the population mode is given by

$$t_p = \tilde{y}\frac{\tilde{x}}{\tilde{X}} \qquad (2.15)$$

The expression of Mean square errors of the naïve product estimator $t_p$'s is given by

$$MSE(t_p) \approx V(\tilde{y}) + \tilde{R}^2 V(\tilde{x}) + 2\tilde{R}\, Cov(\tilde{y},\tilde{x}) \qquad (2.16)$$

Where $\tilde{R} = \frac{\tilde{Y}}{\tilde{X}}$, is the ratio of the population modes of the study variable and the auxiliary variable.

### 3. The Proposed Estimators

In this section we proposed a transformed naïve ratio and product based estimators of the population mode $\tilde{Y}$ as:

$$T_R = \tilde{y}\left(\frac{\tilde{X}+L_1}{\tilde{x}+L_1}\right) \qquad (3.1)$$

and

$$T_P = \tilde{y}\left(\frac{\tilde{x}+K_1}{\tilde{X}+K_1}\right) \qquad (3.2)$$

Where $L_1$ and $K_1$ are suitably chosen characterizing scalar. The bias and mean square errors of $T_R$ and $T_P$ to the first order of approximation, are given by:

$$Bias(T_R) = \tilde{Y}\left[\frac{V(\tilde{x})}{(\tilde{X}+L_1)^2} - \frac{Cov(\tilde{y},\tilde{x})}{\tilde{Y}(\tilde{X}+L_1)}\right] \qquad (3.3)$$

$$MSE(T_R) = V(\tilde{y}) + \left(\frac{\tilde{Y}}{\tilde{X}+L_1}\right)^2 V(\tilde{x}) - 2\left(\frac{\tilde{Y}}{\tilde{X}+L_1}\right) Cov(\tilde{y}, \tilde{x}) \qquad (3.4)$$

$$Bias(T_P) = \tilde{Y}\left[\frac{V(\tilde{x})}{(\tilde{X}+K_1)^2} + \frac{Cov(\tilde{y},\tilde{x})}{\tilde{Y}(\tilde{X}+K_1)}\right] \qquad (3.5)$$

$$MSE(T_P) = V(\tilde{y}) + \left(\frac{\tilde{Y}}{\tilde{X}+K_1}\right)^2 V(\tilde{x}) + 2\left(\frac{\tilde{Y}}{\tilde{X}+K_1}\right) Cov(\tilde{y}, \tilde{x}) \qquad (3.6)$$

Using the concept of maxima and minima we find the optimum value of $L_1$ from the equation (3.4) and value of $K_1$ from the equation (3.5) i.e.

$$L_{1\,opt.} = \tilde{Y}\left(\frac{V(\tilde{x})}{Cov(\tilde{y},\tilde{x})}\right) - \tilde{X} \qquad (3.7)$$

$$K_{1\,opt.} = -\tilde{Y}\left(\frac{V(\tilde{x})}{Cov(\tilde{y},\tilde{x})}\right) - \tilde{X} \qquad (3.8)$$

The minimum mean square errors of the ratio estimator $T_R$ and product estimator $T_P$ for the optimum value of the characterizing scalars $L_1$ and $K_1$ is given by

$$MSE(T_R)_{opt.} = V(\tilde{y})(1 - \rho_{\tilde{y}\tilde{x}}^2) \qquad (3.9)$$

$$MSE(T_P)_{opt.} = V(\tilde{y})(1 - \rho_{\tilde{y}\tilde{x}}^2) \qquad (3.10)$$

**4. Efficiency Comparison**

In this section, we find the condition for which the proposed transformed naïve ratio based estimator of the population mode will have minimum mean square errors as compared to the naïve estimator and naïve ratio estimator for estimating the population mode.

**4.1 Comparison with naïve estimator of population mode**

Using the expressions (2.7) and (3.4), The proposed estimator $T_R$ will be more efficient than the naïve estimator $\tilde{y}$ if,

$$MSE(T_R) \leq MSE(\tilde{y})$$

$$V(\tilde{y}) + \left(\frac{\tilde{Y}}{\tilde{X}+L_1}\right)^2 V(\tilde{x}) - 2\left(\frac{\tilde{Y}}{\tilde{X}+L_1}\right) Cov(\tilde{x}, \tilde{y}) \leq MSE(\tilde{y})$$

$$\left(\frac{\tilde{Y}}{\tilde{X}+L_1}\right)^2 V(\tilde{x}) - 2\left(\frac{\tilde{Y}}{\tilde{X}+L_1}\right) Cov(\tilde{x}, \tilde{y}) \leq 0$$

$$\rho_{\tilde{y}\tilde{x}} \geq \frac{1}{2}\left(\frac{\tilde{X}}{\tilde{X}+L_1}\right)\frac{C_{\tilde{x}}}{C_{\tilde{y}}} \qquad (4.1)$$

## 4.2 Comparison with naïve ratio estimator of population mode

Using the expressions (2.14) and (3.4), The proposed estimator $T_R$ will be more efficient than the naïve ratio estimator $t_r$ if,

$$MSE(T_R) \leq MSE(t_r)$$

$$V(\tilde{y}) + \left(\frac{\tilde{Y}}{\tilde{X}+L_1}\right)^2 V(\tilde{x}) - 2\left(\frac{\tilde{Y}}{\tilde{X}+L_1}\right) Cov(\tilde{x},\tilde{y}) \leq V(\tilde{y}) + \tilde{R}^2 V(\tilde{x}) - 2\tilde{R} Cov(\tilde{x},\tilde{y})$$

$$\left(\frac{\tilde{Y}}{\tilde{X}+L_1}\right)^2 V(\tilde{x}) - 2\left(\frac{\tilde{Y}}{\tilde{X}+L_1}\right) Cov(\tilde{x},\tilde{y}) \leq \tilde{R}^2 V(\tilde{x}) - 2\tilde{R} Cov(\tilde{x},\tilde{y})$$

$$V(\tilde{x})\left\{\left(\frac{\tilde{Y}}{\tilde{X}+L_1}\right) + \tilde{R}\right\} \leq 2 Cov(\tilde{x},\tilde{y})$$

$$\rho_{\tilde{y}\tilde{x}} \geq \frac{1}{2}\left\{\left(\frac{\tilde{X}}{\tilde{X}+L_1}\right) + 1\right\}\frac{C_{\tilde{x}}}{C_{\tilde{y}}} \qquad (4.2)$$

**Remark 4.1.** We know that the naïve ratio estimator is more efficient than the naïve estimator of population mode following the SRSWOR if $\rho_{\tilde{y}\tilde{x}} \geq \frac{1}{2}\frac{C_{\tilde{x}}}{C_{\tilde{y}}}$. Similar results are given for which proposed transformed naïve ratio type estimator is more efficient than

(i) Naïve estimator of population mode if $\rho_{\tilde{y}\tilde{x}} \geq \frac{1}{2}\left(\frac{\tilde{X}}{\tilde{X}+L_1}\right)\frac{C_{\tilde{x}}}{C_{\tilde{y}}}$

(ii) Naïve ratio type estimator of population mode if $\rho_{\tilde{y}\tilde{x}} \geq \frac{1}{2}\left\{\left(\frac{\tilde{X}}{\tilde{X}+L_1}\right) + 1\right\}\frac{C_{\tilde{x}}}{C_{\tilde{y}}}$

## 5. Simulation Study

In this section, we do analysis with a real data set and artificially generated data set as a population. The real data set represents the two different measurements of stiffness, 'Shock' as auxiliary variable $'x'$ and 'Vibration' as study variable $'y'$ of each of 30 boards. The first measurement (Shock) involves sending a shock wave down the board and the second measurement (Vibration) is determined while vibrating the board. The data set was originally reported by William Galligan, and it has also been reported in Johnson and Wichern (1992). We generate the artificial data by assuming the size $N = 5000$ of independent Gamma variable $x_i \sim G(10.00, 0.667)$ and we use the linear relation $y_i = 0.75 + 0.87 x_i + 0.5 z_i$, where $z_i \sim N(0,1)$ for generating another variable for the study purpose with the help of $R$ 3.4.0 software. Given data is ordered pairs of $(y, x)$ and fitted a Gamma distribution for each variable. The fitted distribution of the study variable and auxiliary variable is given in **Figure 1** for the real and artificially generated data sets. We find the parametric estimates of Gamma distribution for the study variable and the auxiliary variable. We also find various descriptive parameters of the study variable and the auxiliary variable listed in the **Table 1** for the real and artificially generated data sets. **Figure 2** shows another graphical representation of the population given in the form of scatter diagram and box plots. The correlation coefficient between study variable and auxiliary variable is $\rho_{yx} = 0.914$ for the real data set and $\rho_{yx} = 0.998$ for the artificially generated data set, which is quite good for our study. We also observe that the value of $P_{11} = 0.4333$ for real data set and $P_{11} = 0.4902$ for artificially generated data set. The correlation coefficient between sample modes is, $\rho_{\tilde{y}\tilde{x}} = 0.5466$ for real data set and $\rho_{\tilde{y}\tilde{x}} = 0.9149$ for artificially generated data set.

The relative efficiency of proposed transformed naïve ratio based estimator and naïve ratio estimator to the naïve estimator of population mode is calculated for real data and is given as:

$$RE(T_R) = \frac{V(\tilde{y})}{MSE(T_R)} \times 100\% \quad \text{and} \quad RE(t_r) = \frac{V(\tilde{y})}{MSE(t_r)} \times 100\%.$$

In our case the subsequent sampling scheme is simple random sampling without replacement(SRSWOR) so there is possible number of samples are $(NCn)$ which are enormously large for both the data sets. For studying all the samples is quite tedious job so we conduct Monte-Carlo simulation studies. In the simulation study we simulated $M = 10000$ samples each of size $n$ and then compute simulated mean square errors, simulated absolute relative bias, simulated relative efficiencies and ratios of 'exact expression of mean square errors' to the 'simulated mean square errors' for the optimum value of the characterizing scalar $L_1$ and we also calculate simulated mean square errors and exact mean square errors for the particular range of characterizing scalar $L_1$.

The simulated mean square errors of the proposed transformed naïve ratio based estimator, naïve ratio estimator and naïve estimator of the population mode given as:

$$MSE(\tilde{y}) = \frac{1}{M}\sum_{i=1}^{M}(\tilde{y}_{|k} - \tilde{Y})^2, \qquad MSE(t_r) = \frac{1}{M}\sum_{i=1}^{M}(t_{r|k} - \tilde{Y})^2 \qquad \text{and}$$

$$MSE(T_R) = \frac{1}{M}\sum_{i=1}^{M}(T_{R|k} - \tilde{Y})^2$$

The simulated relative efficiency of the proposed transformed naïve ratio based estimator, naïve ratio estimator and naïve estimator with respect to the naïve estimator of the population mode for both the data set are given as:

$$RE(\tilde{y}) = \frac{\sum_{k=1}^{M}(\tilde{y}_{|k}-\tilde{Y})^2}{\sum_{k=1}^{M}(\tilde{y}_{|k}-\tilde{Y})^2} \times 100\%, \qquad RE(t_r) = \frac{\sum_{k=1}^{M}(\tilde{y}_{|k}-\tilde{Y})^2}{\sum_{k=1}^{M}(t_{r|k}-\tilde{Y})^2} \times 100\% \quad \text{and}$$

$$RE(T_R) = \frac{\sum_{k=1}^{M}(\tilde{y}_{|k}-\tilde{Y})^2}{\sum_{k=1}^{M}(T_{R|k}-\tilde{Y})^2} \times 100\%$$

We also calculated the simulated absolute relative biases of the estimators $(T_R, t_r, \tilde{y})$ for both the data set are given as:

$$ARB(\tilde{y}) = \frac{\left|\frac{1}{M}\sum_{i=1}^{M}(\tilde{y}_{|k}-\tilde{Y})\right|}{\tilde{Y}}, \quad ARB(t_r) = \frac{\left|\frac{1}{M}\sum_{i=1}^{M}(t_{r|k}-\tilde{Y})\right|}{\tilde{Y}} \quad \text{and} \quad ARB(T_R) = \frac{\left|\frac{1}{M}\sum_{i=1}^{M}(T_{R|k}-\tilde{Y})\right|}{\tilde{Y}}$$

For the investigations of exact mean square errors are how distant from the simulated mean square errors, we computed the following three ratios for generated data set are given as:

$$R(1) = \frac{V(\tilde{y})}{\frac{1}{M}\sum_{i=1}^{M}(\tilde{y}_{|k}-\tilde{Y})^2}, \quad R(2) = \frac{MSE(t_r)}{\frac{1}{M}\sum_{i=1}^{M}(t_{r|k}-\tilde{Y})^2} \quad \text{and} \quad R(3) = \frac{MSE(T_R)}{\frac{1}{M}\sum_{i=1}^{M}(T_{R|k}-\tilde{Y})^2}$$

We obtained $\tilde{y}_{|k}$, $t_{r|k}$ and $T_{R|k}$ viz. Naïve sample mode estimate, naïve ratio estimate, and proposed transformed naïve ratio based estimate from the $K^{th}$ sample for $k = 1,2,...M$.

The simulated mean square errors, simulated absolute relative bias, simulated relative efficiencies and ratios of 'exact mean square errors' to the 'simulated mean square errors' for the optimum value of the characterizing scalar $L_1$ for the various values of the sample size ($n = 51,101,151,201,251,301$) are listed in the **Table 2** for artificially generated data set. The **Figure 3** is the graphical representation of simulated mean square errors, absolute relative bias and ratios with respect to the various sample size ($n = 51,101,151,201,251,301$) for the artificially generated data set. The simulated mean square errors, simulated absolute relative bias, simulated relative efficiencies for the optimum value of the characterizing scalar $L_1$ for the various values of the sample size ($n = 4,8,12,16,20,22$) are listed in the **Table 3** for real data set. The exact mean square errors, absolute relative bias, relative efficiencies for the optimum value of the characterizing scalar $L_1$ for the various values of the sample size ($n = 4,8,12,16,20,22$) are listed in the **Table 4** for real data set. The **Figure 4** is the graphical representation **Table 3 and Table 4.**

We observed that absolute relative bias is very less and close to zero. Maximum absolute relative bias for the artificially generated data set is 0.08% and for the real data set is 2.88%. According to Cochran (1963) relative bias up to 10% is acceptable so in our case it is negligible for the proposed estimator. We observed that ratios of the exact mean square errors to the simulated mean square errors are closed to the one form the Table 2 which indicate that simulated mean square errors is approximately close to exact mean square errors. In the pragmatic way we can use exact mean square errors as simulated mean square errors

We computed exact expressions of mean square errors and simulated mean squared errors for some defined range of characterizing scalar $L_1$ for the real and generated data sets, where the sample size $n = 12$ for the real data set and $n = 151$ is for generated data set. Numerical values are listed in the **Table 5** and graphical representation is in; **Figure 5**. The relative efficiency of proposed transformed naïve ratio based estimator is more and the mean square errors are minimum in the contrast of other existing estimators. The most important thing we observed that exact relative efficiency are independent of sample size but in the other hand the simulated relative efficiency, absolute relative biases depends upon the sample size.

**Table 1:** The summary of the data for real as well as generated data sets.

| | For Real Data | | | | | |
|---|---|---|---|---|---|---|
| Variable | Min. | Lower Qu. | Median | Mean | Upper Qu. | Max. |
| y | 1170 | 1596 | 1680 | 1750 | 1889 | 2794 |
| x | 1325 | 1715 | 1863 | 1906 | 2057 | 2983 |
| | For Generated Data | | | | | |
| Variable | Min. | Lower Qu. | Median | Mean | Upper Qu. | Max. |
| Y | 1.478 | 3.701 | 4.618 | 4.816 | 5.718 | 12.266 |
| X | 0.6444 | 3.4037 | 4.4429 | 4.6724 | 5.6993 | 13.1183 |

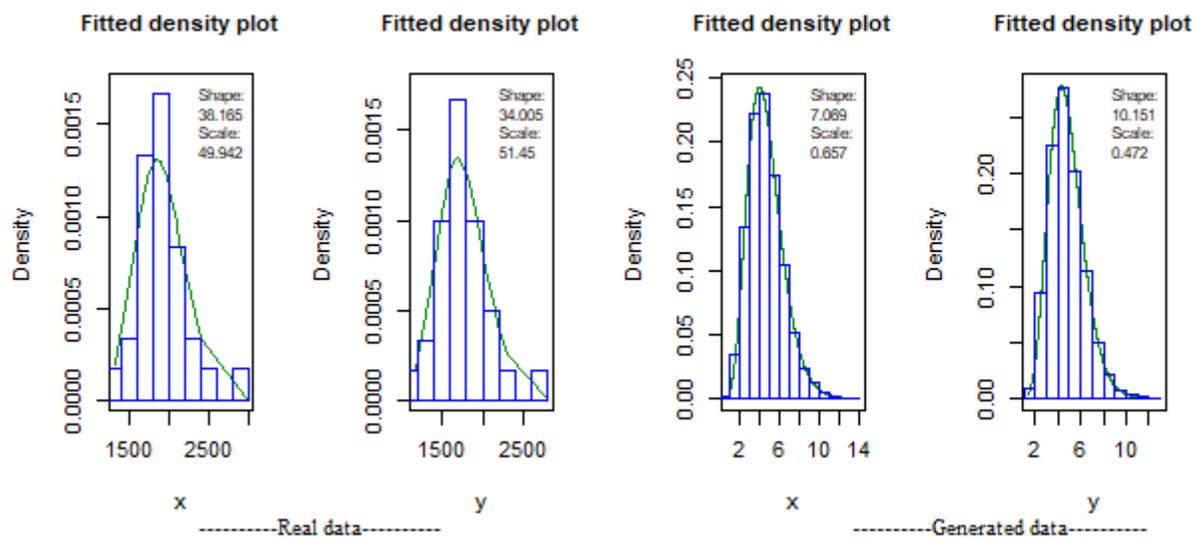

**Figure 1.** Density plot over the histogram of population values of the study and auxiliary variables for the real and generated data sets.

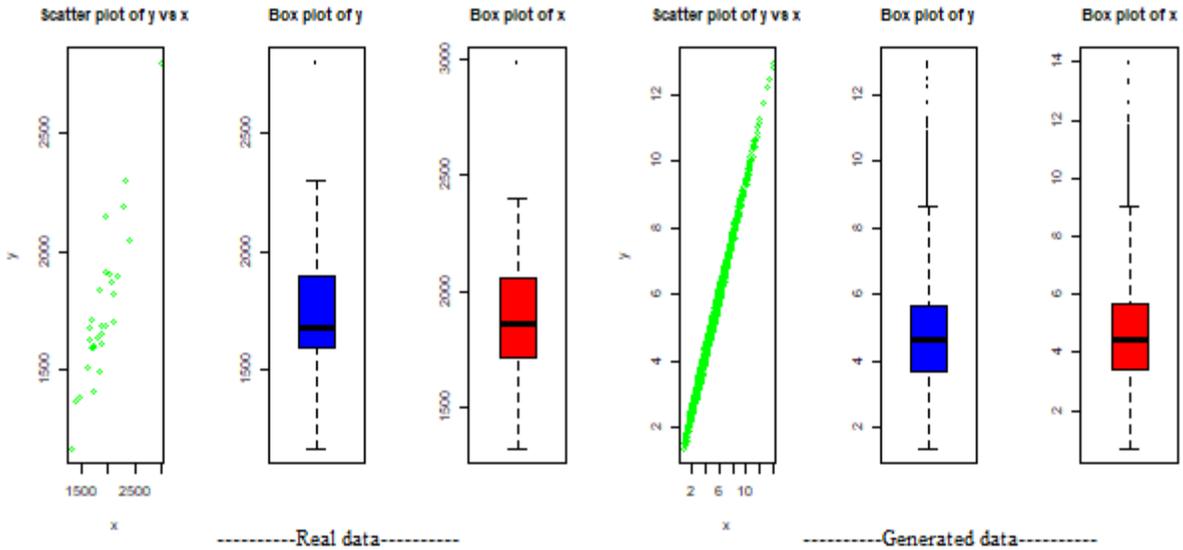

**Figure 2.** Scatter plot of study variable Vs auxiliary variable and Box plot of the study variable and auxiliary variable for Real as well as Generated data.

**Table 2:** Simulated relative efficiency, simulated mean square errors, absolute relative bias and ratios of simulated mean square errors to the exact mean square errors for generated data.

| | For Generated Data | | | | | |
|---|---|---|---|---|---|---|
| $n$ | 51 | 101 | 151 | 201 | 251 | 301 |
| $RE(\tilde{y})$ | 100.00 | 100.00 | 100.00 | 100.00 | 100.00 | 100.00 |
| $MSE(\tilde{y})$ | 0.2515 | 0.1275 | 0.0842 | 0.0637 | 0.0500 | 0.0405 |
| $ARB(\tilde{y})$ | 0.0063 | 0.0000 | 0.0002 | 0.0016 | 0.0006 | 0.0003 |
| $R(\tilde{y})$ | 1.0576 | 1.0427 | 1.0454 | 1.0273 | 1.0372 | 1.0565 |
| $RE(t_r)$ | 414.3328 | 366.3793 | 369.2982 | 374.7059 | 349.6503 | 331.9672 |
| $MSE(t_r)$ | 0.0607 | 0.0348 | 0.0228 | 0.0170 | 0.0143 | 0.0122 |
| $ARB(t_r)$ | 0.0037 | 0.0038 | 0.0017 | 0.0019 | 0.0016 | 0.0015 |
| $R(t_r)$ | 1.1374 | 0.9917 | 1.0021 | 0.9992 | 0.9413 | 0.9104 |
| $RE(T_R)$ | 733.2362 | 604.2654 | 592.9577 | 612.5000 | 574.7126 | 578.5714 |
| $MSE(T_R)$ | 0.0343 | 0.0211 | 0.0142 | 0.0104 | 0.0087 | 0.0070 |
| $ARB(T_R)$ | 0.0003 | 0.0008 | 0.0000 | 0.0008 | 0.0003 | 0.0005 |
| $R(T_R)$ | 1.2636 | 1.0268 | 1.0101 | 1.0254 | 0.9714 | 0.9961 |

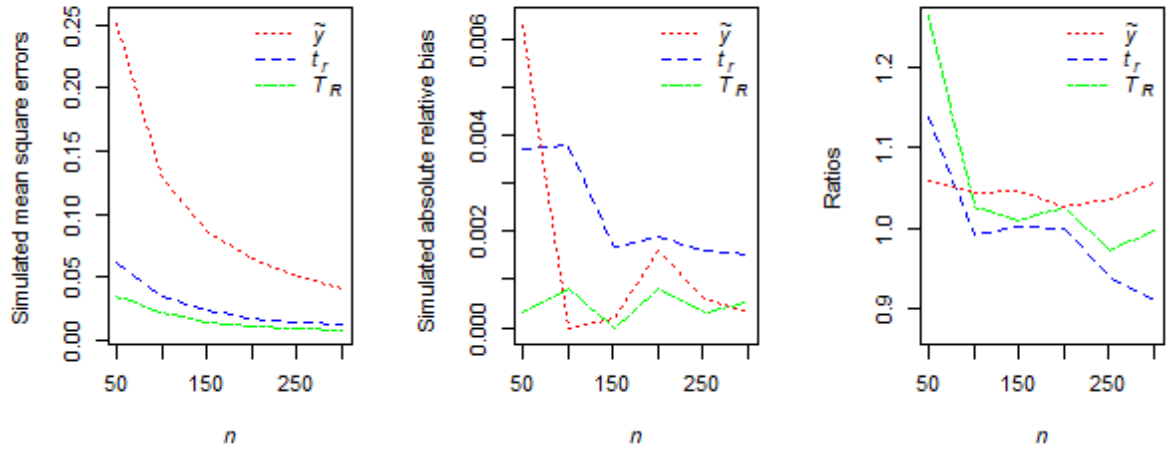

Figure 3: Simulated mean square errors, simulated relative bias and ratios for the generated data.

Table 3: Simulated relative efficiency, simulated mean square errors, simulated absolute relative bias for the real data set.

| | For Real Data | | | | | |
|---|---|---|---|---|---|---|
| $n$ | 5 | 8 | 12 | 16 | 20 | 22 |
| $RE(\tilde{y})$ | 100.00 | 100.00 | 100.00 | 100.00 | 100.00 | 100.00 |
| $MSE(\tilde{y})$ | 62625.3395 | 26749.9930 | 15596.9699 | 9040.5768 | 4589.3017 | 3116.9566 |
| $ARB(\tilde{y})$ | 0.0416 | 0.0320 | 0.0108 | 0.0036 | 0.0068 | 0.0103 |
| $RE(t_r)$ | 121.3924 | 127.1424 | 121.0312 | 119.9680 | 116.6193 | 107.8786 |
| $MSE(t_r)$ | 51589.1676 | 21039.3913 | 12886.7393 | 7535.8251 | 3935.2853 | 2889.3199 |
| $ARB(t_r)$ | 0.0286 | 0.0241 | 0.0053 | 0.0067 | 0.0159 | 0.0189 |
| $RE(T_R)$ | 145.1151 | 150.0710 | 150.2963 | 139.9340 | 133.9168 | 125.6464 |
| $MSE(T_R)$ | 43155.6342 | 17824.8891 | 10377.4810 | 6460.6012 | 3426.9786 | 2480.7377 |
| $ARB(T_R)$ | 0.0288 | 0.0250 | 0.0060 | 0.0035 | 0.0128 | 0.0158 |

Table 4: Exact relative efficiency, exact mean square errors, absolute relative bias for the real data set.

| | For Real Data | | | | | |
|---|---|---|---|---|---|---|
| $n$ | 5 | 8 | 12 | 16 | 20 | 22 |
| $RE(\tilde{y})$ | 100.00 | 100.00 | 100.00 | 100.00 | 100.00 | 100.00 |
| $MSE(\tilde{y})$ | 107655.3 | 59210.41 | 32296.59 | 18839.68 | 10765.53 | 7829.476 |
| $ARB(\tilde{y})$ | 0.00 | 0.00 | 0.00 | 0.00 | 0.00 | 0.00 |
| $RE(t_r)$ | 123.6952 | 123.6952 | 123.6952 | 123.6952 | 123.6952 | 123.6952 |
| $MSE(t_r)$ | 87032.7380 | 47868.0059 | 26109.8214 | 15230.7291 | 8703.2738 | 6329.6537 |
| $ARB(t_r)$ | 0.0130 | 0.0071 | 0.0039 | 0.0023 | 0.0013 | 0.0009 |
| $RE(T_R)$ | 142.5991 | 142.5991 | 142.5991 | 142.5991 | 142.5991 | 142.5991 |
| $MSE(T_R)$ | 75495.0616 | 41522.2839 | 22648.5185 | 13211.6358 | 7549.5062 | 5490.5499 |
| $ARB(T_R)$ | 0.00023 | 0.00021 | 0.00019 | 0.00015 | 0.0001 | 0.0000 |

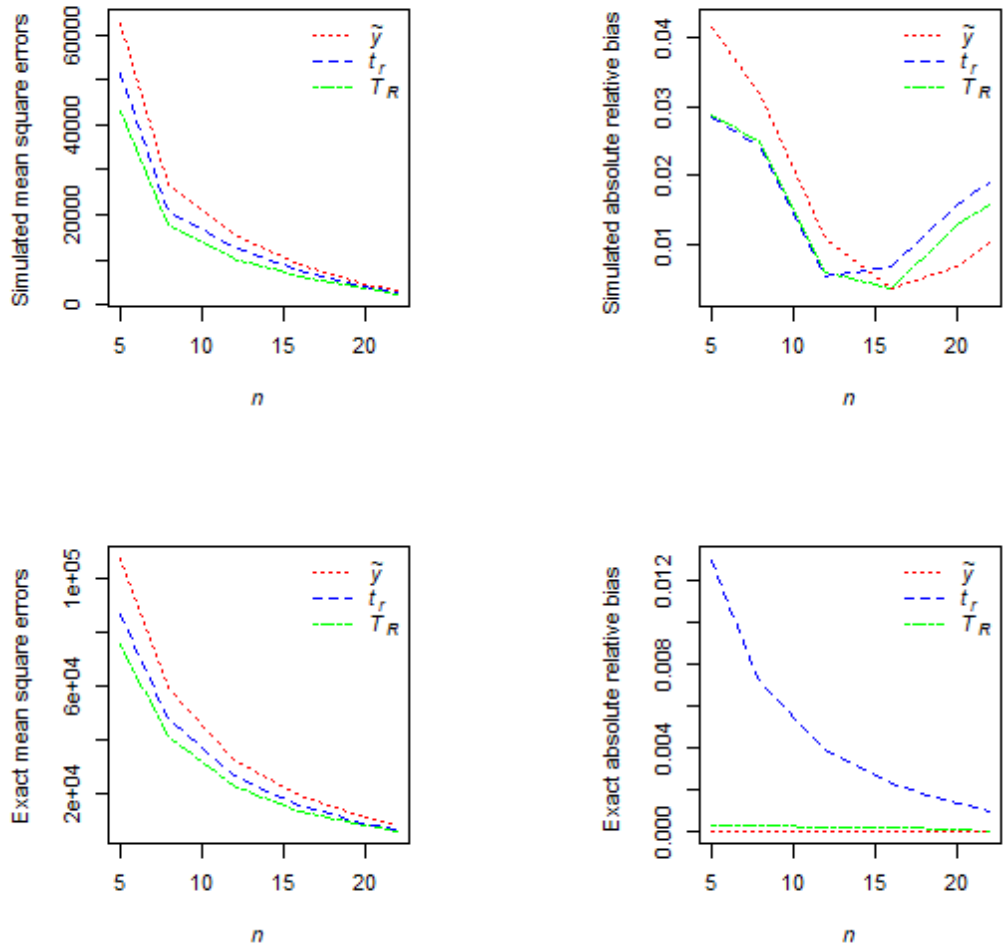

**Figure 4**: Simulated mean square errors, simulated relative bias, exact mean square errors, exact absolute relative bias for the real data set.

**Table 5:** Exact and simulated mean square errors for the different values of $L_1$. The values in (.) and [.] are the existing naïve estimator and naïve ratio estimator of the population mode for real and generated data sets.

| For Real Data: $n = 12$ | | | For Generated Data: $n = 151$ | | |
|---|---|---|---|---|---|
| $L_1$ | Exact MSEs | Simulated MSEs | $L_1$ | Exact MSEs | Simulated MSEs |
| -400 | (32296.59) | (15596.97) | -1.5000 | (0.088) | (0.0842) |
|  | [26109.82] | [12886.74] |  | [0.0228] | [0.0228] |
|  | 33560.94 | 18099.02 |  | 0.1089 | 0.1190 |
| **0.0000** | **26109.82** | **12886.74** | **0.0000** | **0.0228** | **0.0228** |
| 461.5385 | 23348.02 | 10957.48 | 0.3846 | 0.0180 | **0.0179** |
| 923.0769 | 22674.89 | 10422.40 | 0.7692 | 0.0155 | 0.0154 |
| **1064.2364=$L_{1_{opt.}}$** | **22648.52** | **10377.48** | 1.1538 | 0.0145 | 0.0144 |
| 1384.6154 | 22747.60 | 10381.77 | **1.3702=$L_{1_{opt.}}$** | **0.0143** | **0.0142** |
| 1846.1538 | 23097.92 | 10530.26 | 1.5385 | 0.0144 | 0.0143 |
| 2307.6923 | 23542.70 | 10748.97 | 1.9231 | 0.0150 | 0.0148 |
| 2769.2308 | 24005.65 | 10987.78 | 2.3077 | 0.0160 | 0.0158 |
| 3230.7692 | 24454.52 | 11224.85 | 2.6923 | 0.0172 | 0.0170 |
| 3692.3077 | 24876.35 | 11450.78 | 3.0769 | 0.0186 | 0.0183 |
| 4153.8462 | 25266.96 | 11661.96 | 3.4615 | 0.0201 | 0.0197 |
| 4615.3846 | 25626.21 | 11857.48 | 3.8462 | 0.0216 | 0.0212 |
| 5076.9231 | 25955.70 | 12037.72 | 4.2308 | 0.0232 | 0.0227 |
| 5538.4615 | 26257.76 | 12203.62 | 4.6154 | 0.0247 | 0.0242 |
| 6000.0000 | 26534.92 | 13209.29 | 5.0000 | 0.0262 | 0.0256 |

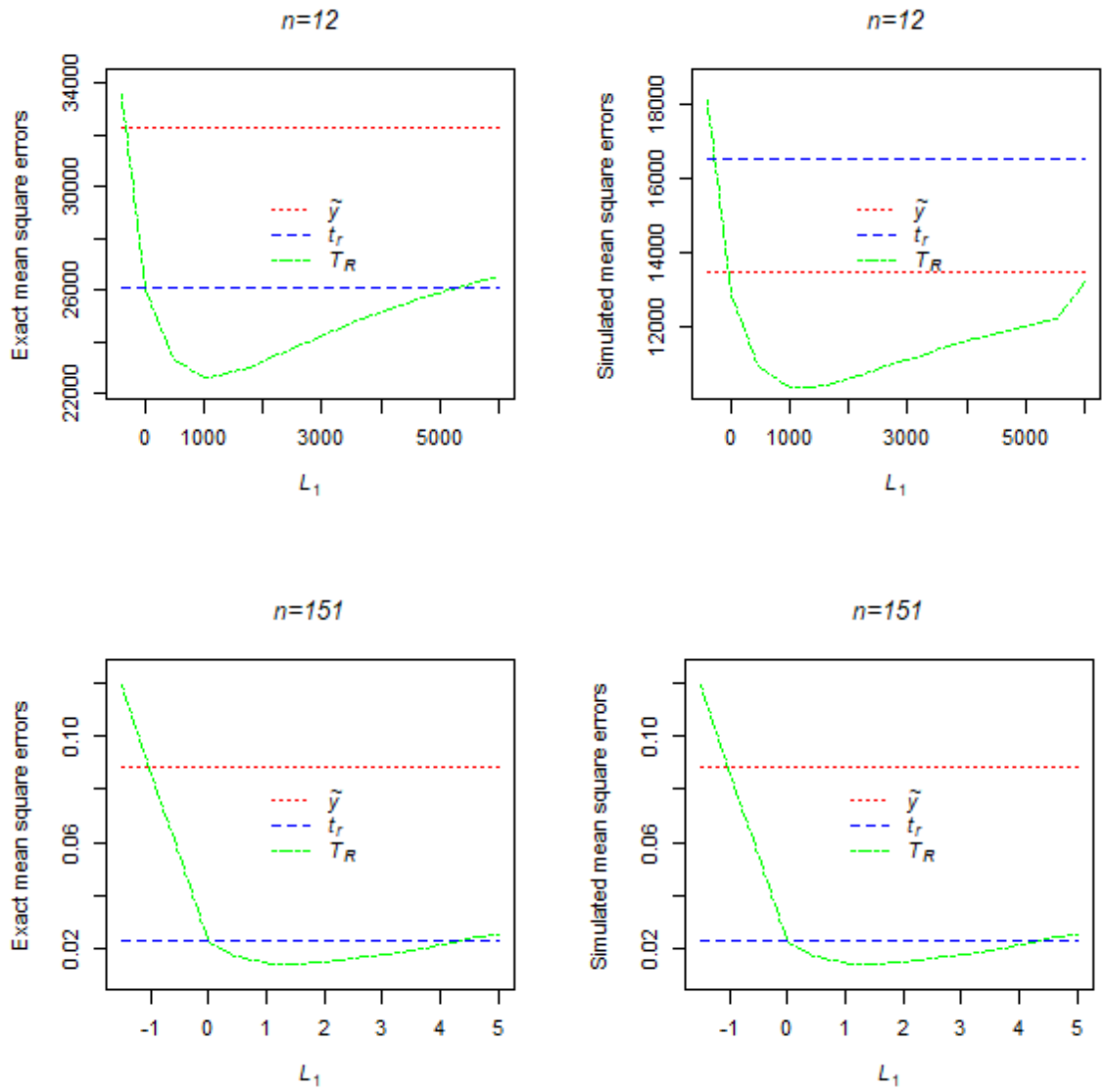

**Figure 5.** The exact and simulated mean square errors for the range of characterizing scalar $L_1$ for real and generated data set.

## 6. Confidence interval

The $100(1 - \alpha)\%$ confidence intervals for the estimators $T_R$, $t_r$ and $\tilde{y}$ are given by

$$T_R \pm t_{(n-1)}(\alpha/2)\sqrt{MSE(T_R)}, \quad t_r \pm t_{(n-1)}(\alpha/2)\sqrt{MSE(t_r)} \quad \text{and} \quad \tilde{y} \pm t_{(n-1)}(\alpha/2)\sqrt{V(\tilde{y})}$$

Where, $t_{(n-1)}(\alpha/2)$ is the value of the $t$-variate at for 95% level of confidence coefficient for the $(n-1)$ degrees of freedom. We calculate 95% exact confidence intervals of the estimates and also calculate simulated confidence interval of the estimates for the various values of sample size of the real and artificially generated data sets. The Exact values of confidence interval and estimated values of the estimators are listed in the **Table 6** for the real as well as the artificially generated data sets. The **Figure 6** shows graphical representation for the real data set and **Figure 7** shows graphical representation for artificially generated data set. The simulated confidence intervals, percent coverage of the estimates, simulated estimates and quartiles of the estimators $(T_R, t_r, \tilde{y})$ are listed in the **Table 7** for the real and generated data sets. The **Figure 8** shows graphical representation for real data set and **Figure 9** shows graphical representation for artificially generated data set. We also draw the violin plots for percent coverage of the estimates in the corresponding simulated confidence intervals for both the data sets. The **Figure 10** shows graphical representation for real data set and **Figure 11** shows graphical representation for artificially generated data set. In the pragmatic way when we increase the sample size the confidence interval of the estimates becomes shorter. The proposed transformed naïve ratio based estimator has shorter confidence interval as well as more percent coverage than other existing naïve estimator of population mode and naïve ratio estimator.

**Table 6:** Exact values of the confidence interval and the estimates of the estimators $(T_R, t_r, \tilde{y})$.

| | $L_{1_{opt}} = 1064.236$ and $\tilde{Y} = 1540.87$ | | | $L_{1_{opt}} = 1.3702$ and $\tilde{Y} = 4.26$ | | |
|---|---|---|---|---|---|---|
| | C.I. | | | C.I. | | |
| Estimates | Lower Limit | Upper Limit | Estimated value | Lower Limit | Upper Limit | Estimated value |
| | | $n = 5$ | | | $n = 51$ | |
| $T_R$ | 1094.55 | 2266.06 | 1680.3 | 4.17 | 4.87 | 4.52 |
| $t_r$ | 1040.72 | 2298.57 | 1669.65 | 3.93 | 4.81 | 4.37 |
| $\tilde{y}$ | 998.92 | 2397.88 | 1698.4 | 4.15 | 5.88 | 5.01 |
| | | $n = 8$ | | | $n = 101$ | |
| $T_R$ | 1252.19 | 2024.31 | 1638.25 | 3.88 | 4.37 | 4.13 |
| $t_r$ | 1197.24 | 2026.26 | 1611.75 | 3.92 | 4.53 | 4.23 |
| $\tilde{y}$ | 1223.49 | 2145.51 | 1684.5 | 3.26 | 4.47 | 3.86 |
| | | $n = 12$ | | | $n = 151$ | |
| $T_R$ | 1318.29 | 1858.83 | 1588.56 | 3.91 | 4.3 | 4.1 |
| $t_r$ | 1288.59 | 1868.97 | 1578.78 | 3.86 | 4.36   | 4.11 |
| $\tilde{y}$ | 1282.42 | 1927.91 | 1605.17 | 3.6 | 4.58 | 4.09 |
|       | $n = 16$ |        |         |      | $n = 201$ |    |
| $T_R$ | 1234.37 | 1637.37 | 1435.87 | 3.87 | 4.21   | 4.04 |
| $t_r$ | 1214.33 | 1647.02 | 1430.67 | 3.76 | 4.19   | 3.97 |
| $\tilde{y}$ | 1204.01 | 1685.24 | 1444.62 | 3.84 | 4.69 | 4.26 |
|       | $n = 20$ |        |         |      | $n = 251$ |    |
| $T_R$ | 1344.58 | 1645.06 | 1494.82 | 4.08 | 4.38   | 4.23 |
| $t_r$ | 1349.29 | 1671.92 | 1510.6  | 4.11 | 4.49   | 4.3  |
| $\tilde{y}$ | 1289.79 | 1648.61 | 1469.2 | 3.67 | 4.42  | 4.05 |
|       | $n = 22$ |        |         |      | $n = 301$ |    |
| $T_R$ | 1438.4  | 1693.41 | 1565.91 | 4.14 | 4.42   | 4.28 |
| $t_r$ | 1406.09 | 1679.89 | 1542.99 | 4.17 | 4.52   | 4.34 |
| $\tilde{y}$ | 1453.47 | 1757.99 | 1605.73 | 3.76 | 4.44 | 4.1  |

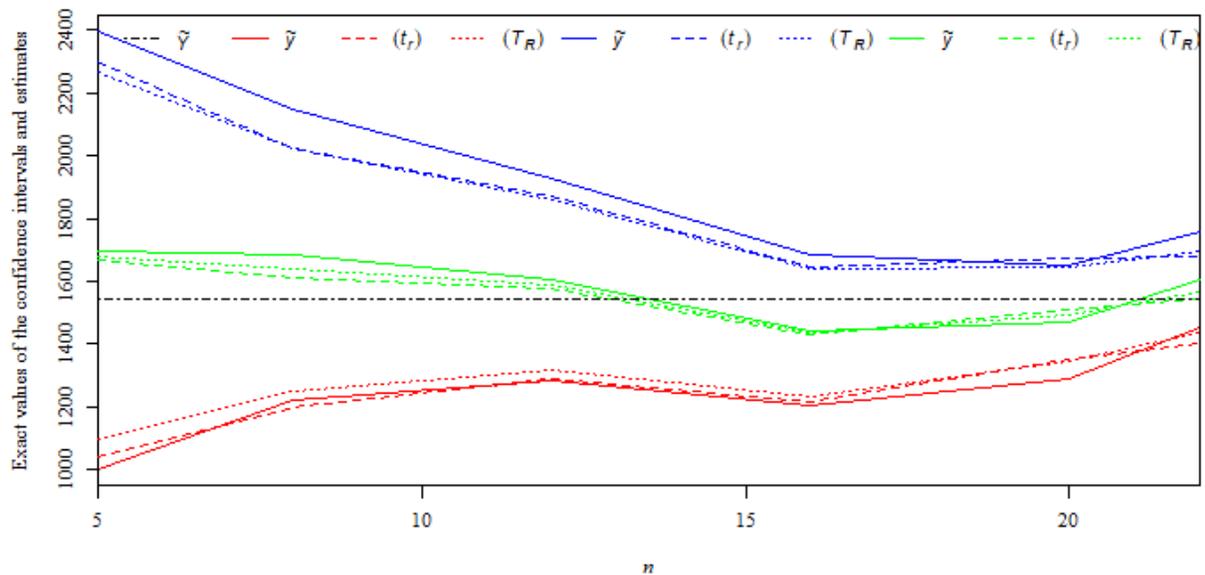

**Figure 6:** Exact values of the confidence intervals and the corresponding estimates of the different estimators for the real data.

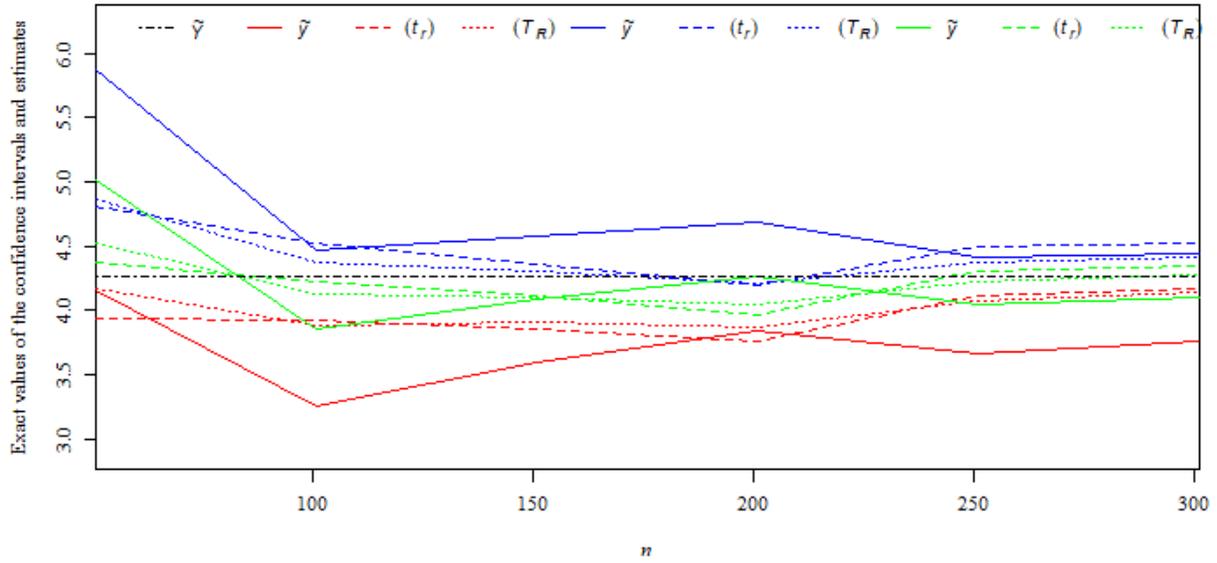

**Figure 7:** Exact values of the confidence intervals and the corresponding estimates of the different estimators for the generated data.

**Table 7:** Simulated confidence intervals, percent coverage of the estimates, simulated estimates and quartile of the estimators $(T_R, t_r, \tilde{y})$

| | For Real Data $L_{1_{opt.}} = 1064.236$ and $\tilde{Y} = 1540.87$ | | | | | | |
|---|---|---|---|---|---|---|---|
| | C.I. | | | | Quartile | | |
| Estimates | Lower Limit | Upper Limit | Coverage (%) | Simulated estimates | Lower Quartile | Median | Upper Quartile |
| | | | $n = 5$ | | | | |
| $T_R$ | 1244.89 | 1925.57 | 95.86 | 1585.23 | 1451.02 | 1579.5 | 1708.7 |
| $t_r$ | 1217.36 | 1952.41 | 92.54 | 1584.88 | 1438 | 1565 | 1716 |
| $\tilde{y}$ | 1201.52 | 2008.43 | 88.97 | 1604.98 | 1466.3 | 1597.2 | 1733.8 |
| | | | $n = 8$ | | | | |
| $T_R$ | 1381.87 | 1777.01 | 94.72 | 1579.44 | 1490 | 1573 | 1657 |
| $t_r$ | 1365.06 | 1791 | 91.97 | 1578.03 | 1481 | 1568 | 1662 |
| $\tilde{y}$ | 1347.04 | 1833.45 | 88.23 | 1590.24 | 1491 | 1593 | 1681 |
| | | | $n = 12$ | | | | |
| $T_R$ | 1406.64 | 1693.58 | 91.76 | 1550.11 | 1479 | 1539 | 1617 |
| $t_r$ | 1389.73 | 1708.32 | 88.99 | 1549.03 | 1468 | 1536 | 1615 |
| $\tilde{y}$ | 1380.76 | 1734.35 | 85.39 | 1557.55 | 1465 | 1556 | 1631 |
| | | | $n = 16$ | | | | |
| $T_R$ | 1426.4 | 1644.51 | 89.07 | 1535.45 | 1482 | 1530 | 1580 |
| $t_r$ | 1413.85 | 1647.26 | 86.33 | 1530.56 | 1474 | 1524 | 1572 |
| $\tilde{y}$ | 1415.27 | 1677.44 | 84.93 | 1547.35 | 1479 | 1544 | 1605 |
| | | | $n = 20$ | | | | |
| $T_R$ | 1440.61 | 1601.67 | 90.97 | 1521.14 | 1484 | 1519 | 1554 |
| $t_r$ | 1430.78 | 1601.93 | 88.97 | 1516.36 | 1478 | 1515 | 1549 |
| $\tilde{y}$ | 1435.3 | 1625.57 | 85.44 | 1530.43 | 1482 | 1523 | 1575 |
| | | | $n = 22$ | | | | |
| $T_R$ | 1396 | 1679 | 97.17 | 1519 | 1491 | 1517 | 1547 |
| $t_r$ | 1378 | 1715 | 95.86 | 1515 | 1487 | 1515 | 1542 |
| $\tilde{y}$ | 1400 | 1695 | 92.21 | 1527 | 1486 | 1523 | 1561 |

For Generated Data $L_{opt} = 1.3702$ and $\tilde{Y} = 4.26$

| Estimates | C.I. | | Coverage (%) | Simulated estimates | Quartile | | |
|---|---|---|---|---|---|---|---|
| | Lower Limit | Upper Limit | | | Lower Quartile | Median | Upper Quartile |
| | | | $n = 51$ | | | | |
| $T_R$ | 4.02 | 4.51 | 99.93 | 4.26 | 4.14 | 4.26 | 4.38 |
| $t_r$ | 3.96 | 4.59 | 98.81 | 4.28 | 4.11 | 4.25 | 4.42 |
| $\tilde{y}$ | 3.62 | 4.96 | 81.84 | 4.29 | 3.95 | 4.28 | 4.62 |
| | | | $n = 101$ | | | | |
| $T_R$ | 4.07 | 4.45 | 99.85 | 4.26 | 4.17 | 4.26 | 4.36 |
| $t_r$ | 4.03 | 4.52 | 98.24 | 4.28 | 4.15 | 4.26 | 4.39 |
| $\tilde{y}$ | 3.79 | 4.73 | 81.6 | 4.26 | 4.02 | 4.27 | 4.49 |
| | | | $n = 151$ | | | | |
| $T_R$ | 4.1 | 4.42 | 99.83 | 4.26 | 4.18 | 4.26 | 4.34 |
| $t_r$ | 4.07 | 4.46 | 98.48 | 4.27 | 4.16 | 4.26 | 4.36 |
| $\tilde{y}$ | 3.88 | 4.64 | 81.56 | 4.26 | 4.06 | 4.26 | 4.46 |
| | | | $n = 201$ | | | | |
| $T_R$ | 4.13 | 4.4 | 99.8 | 4.26 | 4.20 | 4.26 | 4.33 |
| $t_r$ | 4.1 | 4.44 | 98.81 | 4.27 | 4.18 | 4.26 | 4.35 |
| $\tilde{y}$ | 3.94 | 4.6 | 81 | 4.27 | 4.10 | 4.27 | 4.44 |
| | | | $n = 251$ | | | | |
| $T_R$ | 4.14 | 4.38 | 99.71 | 4.26 | 4.20 | 4.26 | 4.32 |
| $t_r$ | 4.11 | 4.42 | 98.23 | 4.27 | 4.18 | 4.26 | 4.35 |
| $\tilde{y}$ | 3.97 | 4.55 | 80.48 | 4.26 | 4.11 | 4.26 | 4.41 |
| | | | $n = 301$ | | | | |
| $T_R$ | 4.15 | 437 | 99.73 | 4.26 | 4.21 | 4.26 | 4.32 |
| $t_r$ | 4.12 | 4.41 | 98.14 | 4.27 | 4.19 | 4.26 | 4.34 |
| $\tilde{y}$ | 4 | 4.52 | 80.92 | 4.26 | 4.13 | 4.27 | 3.39 |

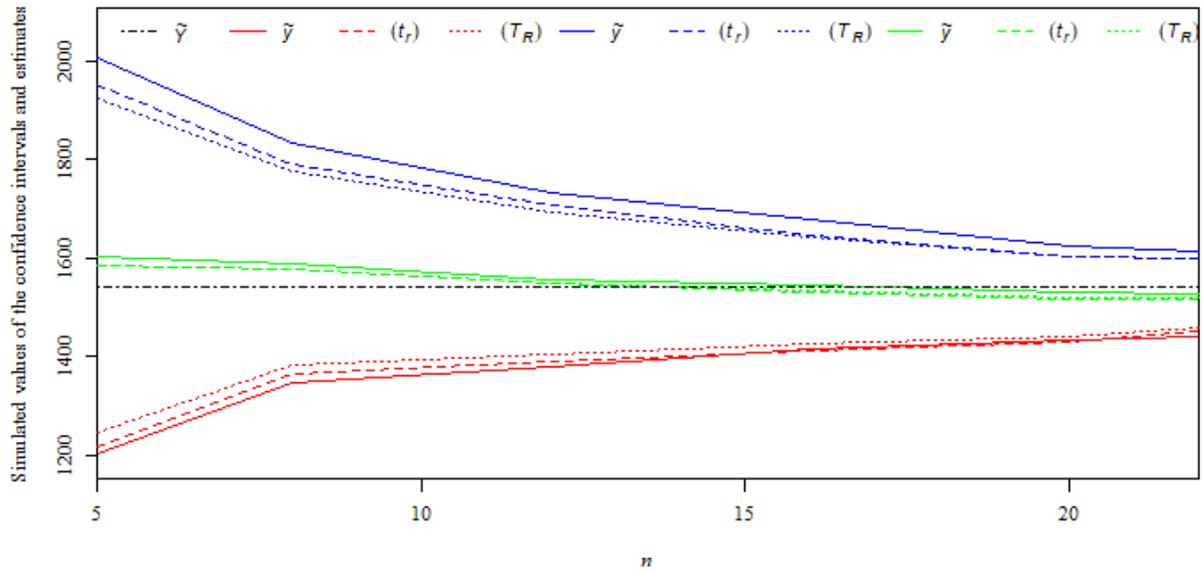

**Figure 8:** Simulated values of the confidence intervals and the corresponding estimates of the different estimators for the real data.

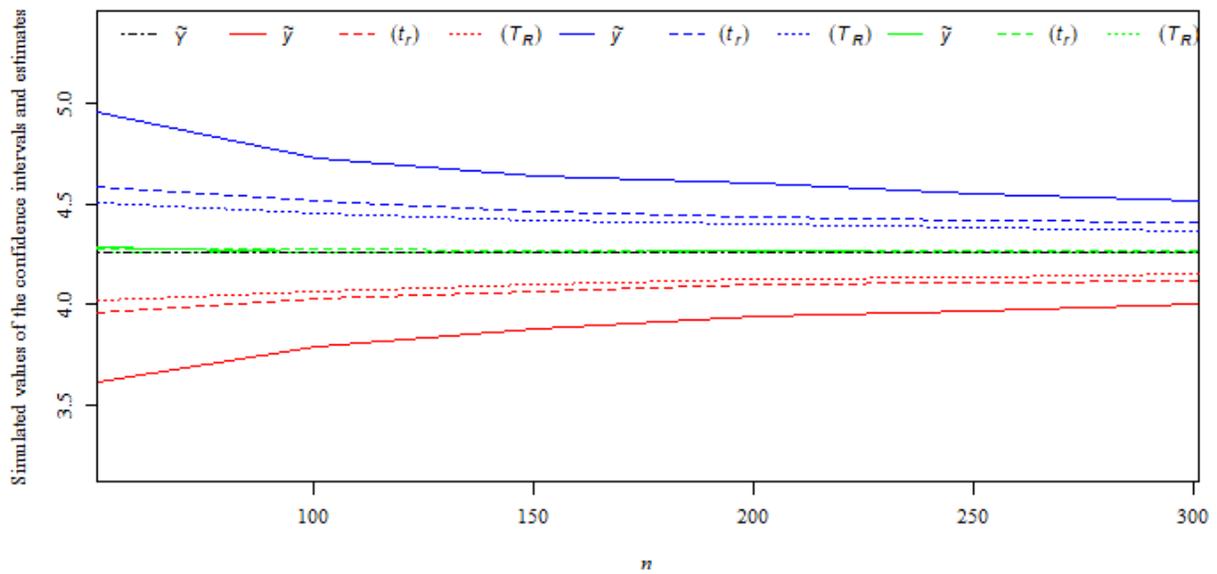

**Figure 9:** Simulated values of the confidence intervals and the corresponding estimates of the different estimators for the generated data.

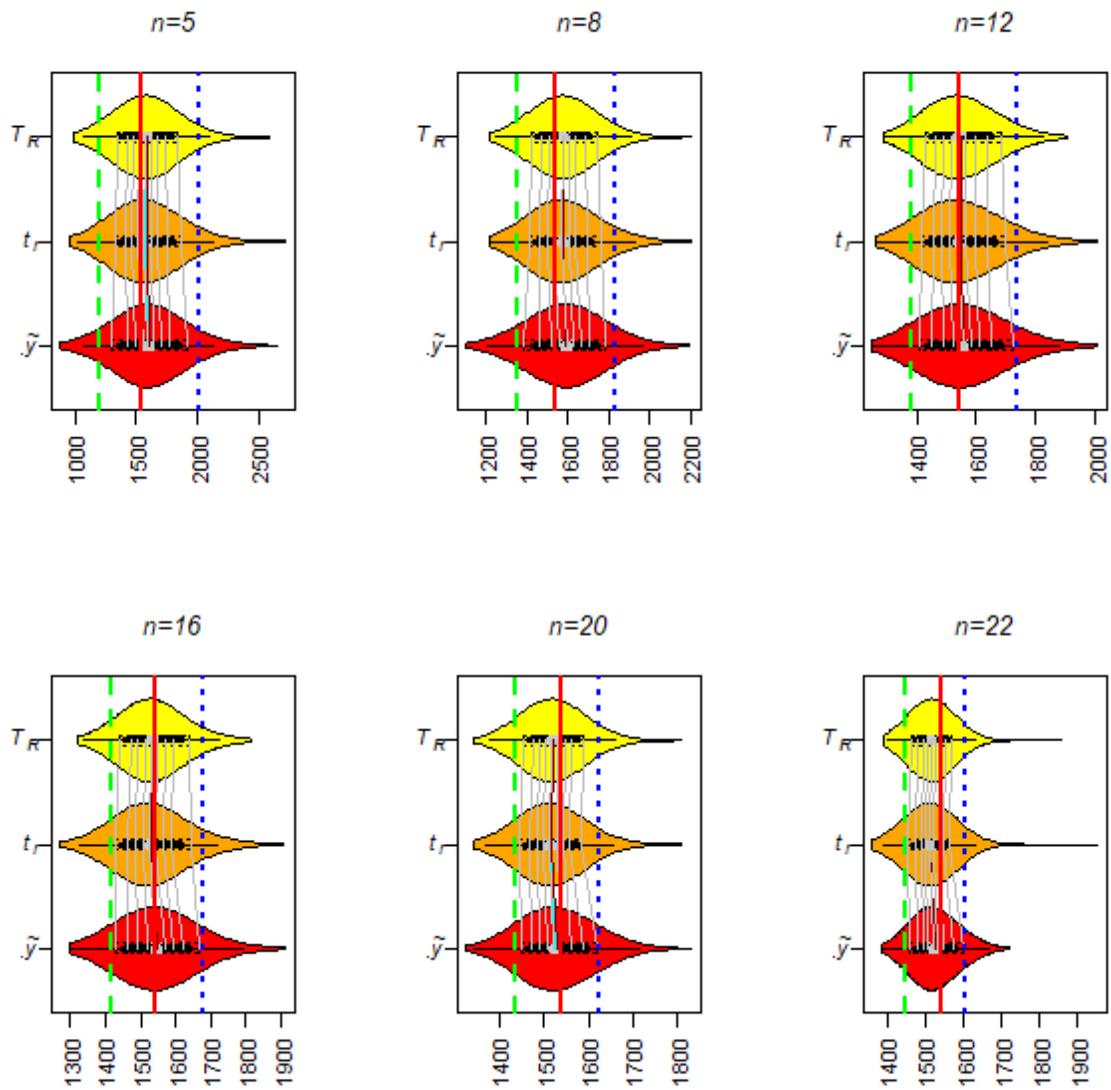

**Figure 10.** Simulated % coverage of the estimates for the optimum value of the characterizing scalar $L_1$ for the real data.

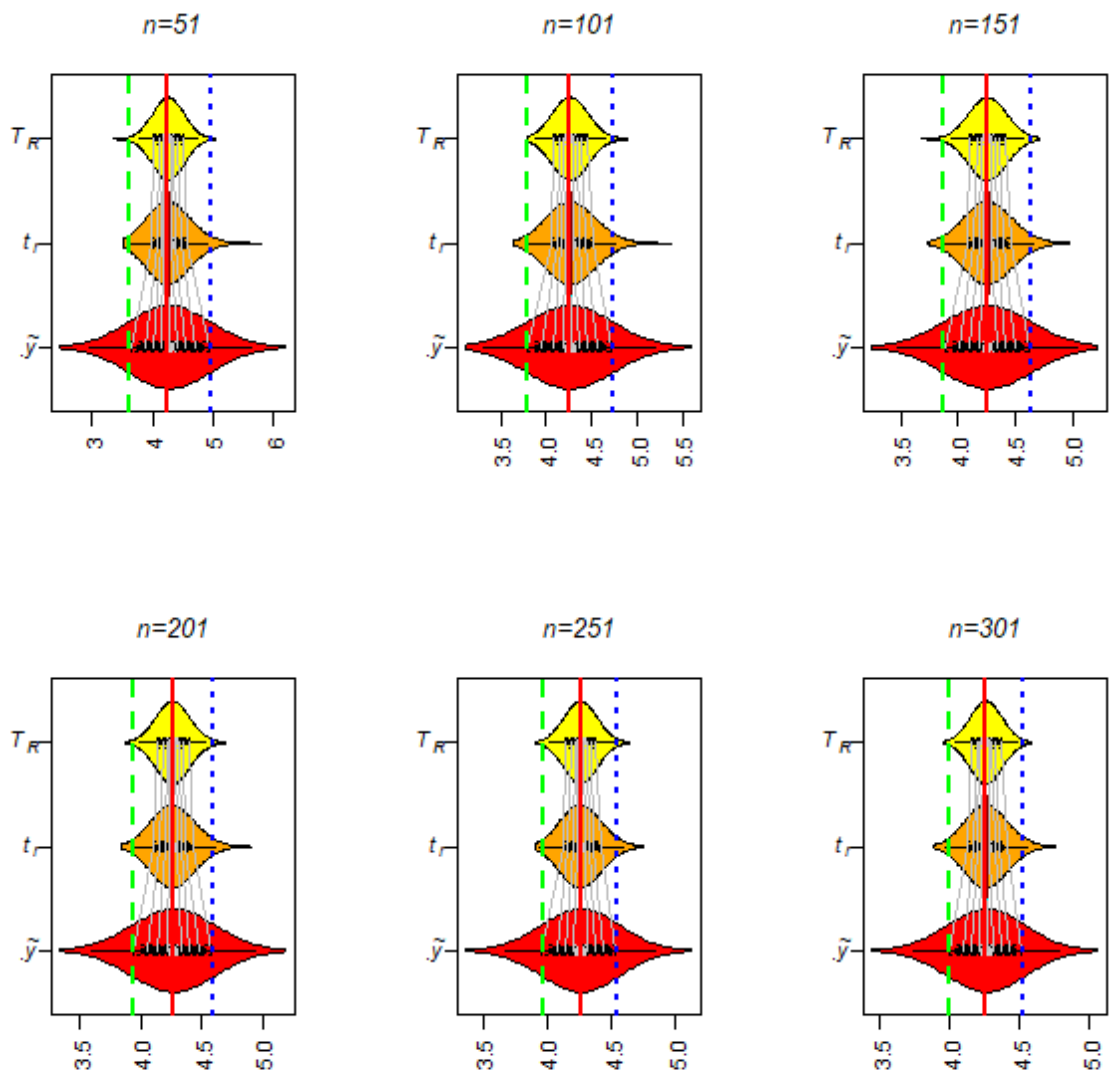

**Figure 11.** Simulated coverage of estimates for the optimum value of $L_1$ for the generated data

## 7. Conclusion

In the present work, we deal with a new transformed naïve ratio and product based estimator using the characterizing scalar in presence of auxiliary information of study variable for estimating the population mode in simple random sampling. The relative efficiency, absolute relative bias, mean square errors, and ratios of the proposed transformed naïve ratio based estimator are obtained and compared for the optimum value of the characterizing scalar to the naïve estimator and naïve ratio estimator of the population mode. We also assessed the performance of the proposed estimator with the certain natural population as well as artificially generated data sets. The percent relative efficiencies of the proposed estimator range between 125.65 to 150.30 for real data set and 574.71 to 733.24 for the generated data set. We observed that a bias and mean square errors of proposed transformed naïve ratio based estimator is minimum in comparison to existing naïve estimator and naïve ratio estimator of the population mode. We also observed that the confidence intervals of the estimates of the proposed transformed naïve ratio based estimator are shorter than existing estimators for various values of sample size. Further, it can also be observed that percentage coverage of simulated confidence intervals of the proposed transformed naïve ratio based estimator is more than existing estimators. So we highly recommend preferring proposed transformed naïve ratio based estimator over existing naïve estimator and naïve ratio estimator for estimating the population mode.

# References

[1] Chernoff, H. (1964). Estimation of the mode. *Annals of the Institute of the Statistical Mathematics* 16:31-41.

[2] Cochran, W. G. (1940). Some properties of estimators based on sampling scheme with varying probabilities. *The Australian Journal of Statistics* 17:22-28.

[3] Cochran, W. G. (1963). *Sampling Techniques*. New York: John Wiley and Sons.

[4] Dalenius, T. (1965). The mode-a neglected parameter. *Journal of the Royal Statistical Society* 128:110-118

[5] Doodson, A.T. (1917). Relation of the mode, median and mean in frequency functions. *Biometrika* 11:425-429.

[6] Grenander, U. (1965). Some direct estimates of the mode. *Annals of Mathematical Statistics* 36:131-138.

[7] Gross, S. T. (1980). Median estimation in sample surveys. *Proceedings of the Survey Research Methods Section (American Statistical Association)*, Houstan, Texas, August 11-14, 1980. Pp. 181-184.

[8] Kendall, M. G., Stuart, A. (1977). *The Advanced Theory of Statistics*. Vol. 1, 4th ed. New York:Hafner Publishing Co.

[9] Kuk, A. Y. C., Mak, T. K. (1989). Median estimation in the presence of auxiliary information. *Journal of the Royal Statistical Society* B 51:261-269.

[10] Robertson, T., Cryer, J.D. (1974). An iterative procedure for estimating the mode. *Journal of the American Statistical Association* 69:1012-1016.

[11] Rueda, M., Arcos, A.,Munoz, J. F., Singh, S. (2007). Quantile estimation in two-phase sampling. *Computational Statistics and Data Analysis* 51(5):2559-2572.

[12] Silverman, B. W. (1986). *Density estimation for statistics and data analysis*. London, UK:Chapman and Hall.

[13] Venter, J. H. (1967). On estimation of the mode. *Annals of Mathematical Statistics* 34:1446-1455.

[14] Yasukawa, K. (1926). On the probable error of the mode of skew frequency distributions. *Biometrika* 18:263-292.

[15] Khare, B.B., Kumar, S. (2009). Transformed two phase sampling ratio and product type estimators for population mean in the presence of non-response. Aligarh Journal of Statistics 29:91-106

[16] Sedory, Stephen A.,Singh Sarjinder (2014). Estimation of Mode Using Auxiliary Information. *Communication in Statistics-Simulation and Computation* 43:2390-2402.